# On the need for a global academic internet platform

Nadja Kutz[*]

March 8, 2008

ACM-Class: K.4.1; K.4.3
Keywords: science, communication, economics, global,
computer-supported collaborative work, transborder data flow, ethics

**Abstract**

The article collects arguments for the necessity of a global academic internet platform, which is organized as a kind of "global scientific parliament". With such a constitution educational and research institutions will have direct means for communicating scientific results, as well as a platform for representing academia and scientific life in the public.

# 1 Academic communication, representation and political processes

## 1.1 Academic communication and its societal representation

In the last years internet communication has taken a leading role in overall societal life. This holds not only true for the western world, but is more and more evident on a global level as well.

New forms of social networking and social communities grew within no time, partially furthered by networking tools, such as wiki's, blogs, cvs repositories, commercial networking sites (e.g. myspace, facebook, xing) or other forms of community forming platforms reaching from online gaming platforms like world of warcraft, over environments such as second life, and online learning platforms to customer services of online stores.

---

[*]email: nadja@daytar.de

Political life has partially merged into this process. Every major political party has at least a website. Political leaders have their own website. International organisations have their websites etc. Political messages are distributed not only via traditional media, like newspapers, TV stations but more and more often via politically colored blogs or directly on media such as youtube. Political communication platforms such as the World Economic Forum [WeFo], Fora.tv etc. provide meeting and information spaces.

However academic life, which had online networking tools long before the internet and whose networking tools (like the html format, server architecture etc) provided the grounds for the current boom takes an astonishingly hidden role in this development.

Universities of course have their own websites. Moreover a great deal of academic life takes place online. Online registrations, augmented learning, student networks, research overviews, publication lists, lecture notes etc are almost standard at every bigger university. Moreover university members take part in investigations or provide information for foundations and political and ecomomic institutions (like the IPCC) and thus they play a strong role in the political communication process. However all these contributions are rather hidden. Even in cases where the participation of academic members is emphasized these are usually mentioned in diffuse terms like "leading climate scientists" or "experts in genetic engineering".

Another important hidden role of academia is the contribution to knowledge accumulation within the internet. This is not only provided via the university portals, but by the participation of university members in collaborative environments such as wikipedia.

In stark contrast to this there is an often strong neglectance of academia and educational institutions in politics. This neglectance takes on various forms. It may be as direct as budget cuts for research and educational institutions or it may be more subtle with methods reaching from restraining the autonomy of universities, interference of politics in academic processes with ideas like "elite formation" to concrete structural desicions like employment and funding regulations.

These political measurements usually take place on a national basis, although research is highly international.

The international organizations which are devoted to represent educational institutions like the UNESCO provide information on educational topics and in part also on research content. They provide tools for collaborations. However, they are mediators and their mediating role is usually limited, which results e. g. in predefined priorities.

Similar things hold true for Science organizations, i. e. they represent scientific life to a certain degree and mediate between academia and society.

This role is important however not exhaustive enough.

In particular the "weakness" of science organisations to represent educational institutions has a structural reason. On one hand it is the relatively small organisatorial size (like the UNESCO Sciences Sector has about 200 staff members (which is small if you are looking for a direct adressee to set up on a science related question) on the other hand it is the very role as a mediator which diminishes the influence of a science organization.

The above should serve as a fast explanation that there is and why there is a certain lack of a direct active representation of academia and academic questions in societal life. Such a representational gap could – at least in part – be filled by an official academic electronic platform, which is directly run by all (or almost all) higher–educational institutions, i. e. universities in the world.

This excludes many good thinkers and artists but considering only university members makes the authentification and organization easier. Last not least the system of universities spans a global net with a rather (emphasis on: rather) high neutrality towards cultural and gender sensibilities, a huge expertise and a good access to local administations.

## 1.2 Academic communication and political processes

Besides possibly filling a representational gap a "science parliament" might emphasize its role as a global consultant. The purpose of this section is firstly to briefly recall the structural sensitivity of political systems and secondly to briefly recall the role of consultants in political systems. A profound political analysis is definitely beyond the scope of this article. The reminder should merely serve as a motivation why a science parliament could act as a consultant.

Democratic systems can be very sensitive to rather subtle organisatorial differences such as between representative democracy, direct democracy, between voting systems, concerning control of power (legislative, executive, jurisdiction) a.s.o. As an example one can compare the Weimar republic and the current german democratic system. It is more or less undisputed that the instability of the Weimar republic were partially due to its democratic organisatorial structures. Of course this has to be seen in context with the historical circumstances, but as a matter of fact the Weimar republic saw 20 cabinet changes in 14 years, whereas the current and former western Bundesrepublik of Germany had 21 changes of government in about 58 years.

Another interesting point when looking at democracies and their representational character is the social and psychological origin of politicians. For most democratic systems the social mixture of politicians does not mirror the social mixture of the corresponding society. A famous example for the case of Germany is the high percentage of lawyers in the german government (which is apparently mostly due to the socalled jurist privilege) [DaBu].

Psychological processes which are involved with raising to and staying in political power within a democratic system are quite complicated. A politician has to be stress resistant, stable or at least emanating stability, resistant to intrigues, be able to make fast and far reaching decisions, which can have vast implications. A politician has to be responsible etc. In short: a politician is a certain kind of a human. This implies that an average of politicians would very probably act quite differently than an average of the overall population, which has its advantages and disadvantages. Group dynamics and for example the confirmation bias whereby we seek and find confirmatory evidence in support of already existing beliefs and ignore or reinterpret disconfirmatory evidence [MSh06] are adding another psychological component.

Another wellknown fact is that the concrete paths politicians choose in their political daily life are often to a great extend informed by consultants and lobbyists, which are mostly representing economic forces.

It makes sense to have consultants – a politician just does not have the time to dig through all the details, which are often needed for a political decision. However as outlined above the choice of consultants seems to be a rather obscure and often quite psychological issue. It is usually not very representative.

This is an obvious violation of the idea of a democracy – given that there is a democracy in a country.

According to the socalled third Transformation Index of the Bertelsmann foundation:

> *Despite the continuing worldwide economic growth of the past few years, mass poverty remains the central problem in most developing countries, and the majority of people have no lasting share in this prosperity. And although the number of governments determined by free elections is growing, many people are still excluded from political decision-making or are actively denied other political and civil rights. This is the sobering conclusion reached by the third Transformation Index (BTI), an international comparative study of 125 developing and transition countries.. . [BTI08]*

Moreover according to Mr. Janning, globalization expert of the Bertelsmann Foundation:

> *"From the global perspective, advancing globalization is producing greater overall growth and prosperity, but not in a fair or sustainable way. The*

> *positive effects of globalization are not benefiting the majority of people and it is not sustainable for the future. The failure, but also the solution to these problems lies in the reform capability at state and government level."*

Summarizing – next to its role of being a representative of the global scientific community, as had been outlined in the preceding section, the science platform could act as a more or less neutrally, transparently, and globally acting consultant. It may saveguard politicians if they have to make controversial desicions, which are facing or will face manhood soon. It could empower the UN to enforce desicions against local warlords. Thus it could change the political landscape without making apriori sensitive structural changes at the political systems themselves, which does not exclude that one keeps thinking about them. Moreover it is in principle possible to use the "science parliament" for informing structural changes such as adaptive management etc.

Nevertheless it should be emphasized again that the main task of such a platform is to serve as an instrument which would allow for a better representation and coordination of global academic knowledge and not as a "shadow government".

# 2 Scientific Methods and the Validation of Scientific Questions

## 2.1 Scientific method, knowledge accumulation

In this section I would like to briefly discuss the role of the scientific method and the validation of scientific questions mainly at the example of math, computer science and physics.

The purpose of this is to explain to a nonscientific audience why the desicion process in science is different from that in politics and society. However the procedure of how "the" scientific method works gives us also indications of how the proposed internet platform may work.

Due to the logical nature of math (the language for physics) the evaluation of a given scientific question or hypothesis is relatively straightforward.

In particular mathematics provides even sometimes notions on wether a question is solvable at all, on how complex a question may be or on how random an answer is.

Mathematical assertions can be *checked for logical consistency*. Interestingly the computer has become more and more important in this in the last years.

Assertions in physics can to a great extend be *checked by measurements and observations*. Physical models/hypothesis/theories (i.e. the mathematical description of physical entities) have to be validated in accordance with these measurements/observations and in accordance with the mathematics describing them.

The whole process must be objective so that the scientist does not bias the interpretation of the results, which includes that the measurements/observations must be in principle reproducable in order to verify them.

In short there is a quite established widely accepted method for checking hypothesis' in mathematics and physics, which applies to a great extend also to other sciences like biology and chemistry and partially also to humanities, like social sciences and also to economics.

The way however how to set up a hypothesis and the question what questions should be asked is usually not straightforward, it is a process which involves imagination, intuition and sometimes also scientific fashions.

## 2.2  Scientific method, review of results

The review/verification process of a result includes various formal steps including the prepublication of results, which reaches from internal discussions with trusted experts to putting them on an electronic archive system as so-called e-prints.

Lately there had been some examples, where – mostly well established faculty members – put drafts of their scientific results out for discussion on websites, as a kind of preprepublication. However this presupposes that the work had reached a certain stage of maturity and that the authors are prepared for discussions.

The archive *arXiv.org*, which was founded in 1991 takes a prominent role in that, i. e. here almost all math and physics publications are freely prepublished and sorted in a content-classification system.

The final step of a publication is then usually done in a peer reviewed journal, where peer review means that the work is independently reviewed by usually at least 2 anonymous experts (the author is usually not anonymous). The anonymity guarantees to a certain degree that the work is investigated solely in terms of content and not in terms of things like personal sympathy. Whereas it should be remarked that is is quite unusual that "negative results", i. e. cases where research lead e.g. to no result are published at all, although the description of these cases could constitute valuable information.

This is (very) roughly what people mean by the scientific method of knowledge acquisition (please see also the wikipedia portal on "scientific method"). In particular this method has been designed to ensure objectivity and designed to avoid in particular biases, like the above mentioned confirmation bias.

But as mentioned before "the" scientific method is not a fixed recipe. It is an ongoing cycle, constantly developing more useful, accurate and comprehensive models, hypothesis and methods.

By the above it is also clear that this system can have failures, in particular the sensitivity of the system to funding and rewards is a delicate issue.

## 2.3 Scientific methods, failures, scientific integrity

The purpose of this section is to decribe the sensitivity of the scientific method with respect to funding in order to provide an insight in possible vulnerabilities of the platform.

The sensitivity of the scientific method to funding starts with the choice of questions. If research has to lead to certain results in a predefined way (like via timelines in a research proposal) then questions will be made with respect to wether this can be achieved at all or not, which implies that questions which are presumably too hard to solve will be left out in such proposals. This holds also true for frequent evaluations, where usually only "positive" and "final" achievements are awarded (which are in terms of science funding often counted as number of publications and number of patents), i. e. again – in terms of evaluation you better choose a subject which has some chance of being successful. Hence the shorter and more limited the proposal/evaluation cycles are, the more results will be "small" results.

Small results or "almost fully satisfying" results can sometimes be useful e. g. in industrial mathematics, where an intelligent mathematical optimization can do sometimes wonders and may already be a sufficient progress considering the invested time and money. But think of how long it took to prove the Fermat conjecture (about 400 years) and imagine how many people would try to apply for a grant proposal in a similar case.

Funding problems can also work as a test case for scientific integrity, i. e. wether the principles of the scientific method are violated. This needs no further explanation – also scientists may be corrupt. However the scientific methods makes corruption much harder then in ordinary life. So funding problems result usually rather in unpleasant interactions among scientists than in wrong assertions. However funding policies may distort the overall picture, like if you look for evidence only in a certain direction then this may lead to insufficient and even wrong conclusions.

Despite the usually high integrity of scientists the problem of scientific integrity has to be mentioned – especially in context of industry/politically funded expertises. The more one individuum or a small research group is dependent on certain funds the higher is the danger of violation of scientific integrity. Likewise this indicates that the more diverse and the higher the number of involved groups in the discussion of a scientific question is, the more integrity can be expected. Furthermore the more open the process of developing a solution is the more peer review will automatically ensure more integrity. Again here the dependence on funding/rewards may result in fears that collegues snatch away intermediate results etc. and thus in hiding the work.

For the case of the electronic platform this implies firstly that the running expenses of the platform has to be made by the universities alone, wether they get reimbursed by an overall higher budget is another question. An initial extra fundraise to install the technics, etc. however may be useful. Moreover not much direct research, which depends on fundings should be involved with the platform (too expensive), but *available information should be rather gathered* for an evaluation process. Secondly, discussions should involve as many work groups as sensible. The more wide-spread and more diverse the groups the more it will also be hard for lobbyists to influence. Scientific discussion should be as open as possible. However it may be necessary to hide work and contributors for preventing lobbyism or for other reasons. Experiences with information blockades after nuclear accidents are an example, where e. g. politics interfered with the pure demand for scientific information. Thirdly the sort of questions to be adressed has to be of public interest, where public may include the scientific public only. Particular benefits of companies have to be avoided or at least discussed openly, as they probably cant be avoided sometimes, but this holds true in general for scientific results.

## 2.4 Further Implications for an electronic platform

There is another aspect one should mention. The scientific method deals with scientific questions. Often the scientific questions to be discussed are in strong relation to e. g. economical, juridicial and ethical questions. A natural-scientific judgement which involved the scientific method may need to be evaluated or juxtaposed in terms of considerations with respect to (economic, political) realizability and ethics. An example: The use of genetically modified plants may impose severe health risks. We may come the point where one has to use genetically modified plants in order to feed the planet. (It is not necessary up to now I think!) So this question has to be discussed in conjunction with these constraints or at least juxtaposed to them. The humanities sections of universities are a very valuable partner in doing this.

This is why the platform could also further interdisciplinarity.

## 2.5 A workflow proposal for the electronic platform

The general schemes of how the scientific method works gives us indications about the possible design of the proposed internet platform/network. The exact technical realization of such a platform is indeed a sensitive issue and beyond the scope of this article. Here a proposal for a general workflow scheme:

### 1. Define questions

The notion of the platform working as a "parliament" means that there are predefined questions. These questions come from society and science itself. Just like laws or societal questions are discussed in a parliament. So society is basically doing the "what-question-should-be-to-be-answered-search" part of the scientific method. The "parliament members" or "experts" are faculty members of universities. The parliament itself is run by universities. It would actually be good if the compilation of questions would be preprocessed and discussed by a forum which is open to everybody, just like say wikipedia.

The questions which are suitable for investigation need to be stated in a precise manner, i. e. scientists may have to reformulate them or dissect them in terms of scientific validation, and economical, political and moral subquestions. The questions relevance has to be established, it has to be ensured that particularism is avoided and it has to be decided wether a question will be made into an official question and as such published on the platform.

### 2. Supply expertise and data

Experts need to supply data to a given official question, which means available scientific material. This presupposes an initial choice of experts, which may supervise the gathering of material and of further experts. Hence this process is similar to the work of an editor of a journal, who assigns communicating faculty and these in the turn assign referees for a work. If global experts are electronically registered and when their expertise is classified via keywords, like in the Mathematics Subject Classification [MSC] then the expert retrieval is fairly simple. However the validation of a question may not necessarily be confined to experts. Non experts could a priori have the possibility to contribute, at least by commenting and providing data. Often e. g. students are very well if not better informed and may contribute at least for data gathering. It should be possible to invite experts from outside the university faculty especially e. g. members of research institutes.

If questions are related to issues of specific nations then it has to be decided wether this is a national or even more local question. If this is not the case it has to be decided wether there has to be a nationality balance in accordance to maybe UN proportions.

### 3. Formulate answers/hypothesis'

Based on their data/expertise experts may formulate answers (hypothesis) and "prepublish" them in a "library" next to a discussion forum corresponding the question or if this is necessary assert that no hypothesis can be made, as e. g. there is not enough material etc. Depending on the question, conferences may be needed (like in the case of the climate change discussions). Here NGOs may play an important role.

### 4. Evaluate answers, publish them

Based on the discussion preliminary or final answers can be formulated and officially published as such. The answers can be explained via the gathered data. The presentation of the results will certainly need a good collaboration with science communicators/journalists in order to avoid communication problems like for example it happened in the report about childhood cancer in the vicinity of nuclear power plants [RaCh08]

In the generic case together with the answer it should be made visible how many experts are in favour for which answer and to what extend. This is what I would call a "vote" or "poll". It may be that experts decide that some aspects are more important than others.

The exact questions of how to poll and decide on the results has to be decided by the experts, however there could be simplified voting mechanisms. In particular simple questions could be decided to be answered via voting on multiple choice questions. This may sound crude especially in context of the careful design of the scientific method, but it may often be sufficient for certain questions or at least for intermediate decisions, like about the issue of relevance etc. For a nice introduction to voting see e. g. the AMS math awareness month website [AMS08]

Depending on the question, the vote as well as the electronic discussion of results itslf must in principle bear the possibility to be made anonymous in order to saveguard the involved scientists as pointed out earlier.

**5. Set timeline**

A timeline for further investigations and validations has to be given next to the "answers". It should be discussed wether and how one could adjust these to strategies like adaptive management.

So in principle the workflow of the platform is not so much different from the scientific day-to-day practise in that its workflow resembles the workflow of the scientific method. However there are differences to the day-to-day practise. I would like to emphasize some of them, as well as emphasize some other relevant points:

- The platform would collect and sort structural data from all universities worldwide and thus provide a worldwide academic network in a electronic-semantically connected way.

- Besides being an organisatorial framework the platform will have a task, namely to represent the global network of universities and to provide answers to scientifically difficult societal questions. The questions will come – at least in part – from society, i. e. in particular the "answering service" by such a platform will be seen as a service of academia to society.

- The amount of answers and the work which will be involved with them is freely adjustable. I. e. if universities dont have enough resources they may decide to terminate the service. Likewise if societies are not content with the service they will proceed in cutting down science budgets.

- The answers to the given questions will in principle already be existing, and not researched i. e. the answers should reflect the current scientific knowledge rather than constitute research. I. e. the main value of the platform is that experts provide and connect information and expertise, rather than that they do research. This doesn't exclude of course that this involves small short term research or that further research may be necessary (see timeline).

- The plattform could be used as a call-in instrument. If scientists are concerned about certain questions, they could call in collegues rather easiliy. Since everything is electronic, these calls can be simply categorized according to relevance, local connectivity etc. Thus mailing lists could be assembled very easily.

- Since the infrastructure of universities is used (computers, rooms), the cost can be kept relatively small.

- Most information which is needed for the platform is already existing, electronically available information (like lists of faculty members, e-prints, open access journals etc.). This information needs to a great extent only be connected. This is more a technical challenge than an organisatorial one. In general, organisatorial regulations should be held minimal and scientists should be trusted in their ability of self-control and self-organisation (given acceptable living and working conditions).

    The expertise of platforms like that of the UNESCO especially with their unitwin networks [UNITWIN] or organisations like sense-about-science [SAS] and other organisations are very valuable and one should think about how to include them into the process.

# 3 Further possible tasks of the platform

The below section is mostly intended to encourage discussion, about the organisatorial structures of economies and their relation to the mechanism of assigning values, which is usually called "pricing".

In particular the topics in this section are intended for thinking about wether the scientific platform could serve as a tool to inform about actual costs and/or indications for pricing.

This section contains a rather simplistic approach to a complicated subject, however simplifying often helps to shed light on the main constituents of a system. On the other hand simplifying to much may rather blur the involved main mechanisms.

In order to explain what is meant by that and what could be the motivation for setting up such a "pricing table", it will be necessary to make a little excursion on what is involved when a value – a price – is assigned to a good. In particular it will be necessary to discuss main forms of values in order to get an understanding of the main mechanisms of pricing and thus serve as an explanation on why the current pricing mechanisms in finance are insufficient (at least in my point of view) and how they could be enhanced by a platform, as proposed above.

## 3.1 Assigning values

In order to study the process of evaluation or pricing one should discuss the notion of a value. For brevity both, the name of a value and its actual quantity will be called a value. It should be clear from the context what is meant.

In the following I will distinguish between three different types of values. I will call them **measurable values**, which are in opposition to rather *emotional values* such as what will be called **subjective values** and **rational subjective values**.

**measurable values**

If you have a sieve and a bowl of sand then the large sand grains stay in the sieve and the small ones drop through, i. e. one can order the grains according to their size and hence the here involved value is called size. Moreover there was no need to invent this value as it was physically already immanent. So a measurement (in the example the measurement of the grain size) is a process where a value is assigned to something, i. e. it is a kind of evaluation. In the case of the measurement this evaluation is done via a physical process (sieving). In the following I would like to call also processes, which can be mathematically quantified, but which depend on a set of values/measurements and the corresponding values as *measured values relative to something*. So e. g. if I have two grains of different weights then I call their compound weight a measured value relative to the given individual weights. (Likewise on could see the grain size as a measured value in relation to the size of the sieve holes). However I do not want to fix this notion too much. The important point here is that for the determination of the value we make use of scientific procedure which is more or less exact repeatable and describable. Nevertheless it should be pointed out that the choice of what and how is also related to attention. I. e. if we stick to the example of the sandgrain then the different sizes of the sand particle caught our attention and made us prepare the experiment with our sand and sieve in this particular way. This is no bad thing per se but it is important when talking about objectivity in scientific reasoning. Good scientific reasoning tries to reduce the attention factor.

**subjective values and rational subjective values**

Another method of assigning values is by a subjective and mostly emotional judgement, like e. g. to assert that one likes one person more than the other or both equal etc. means to assign a *subjective value*. Here it is even harder to seperate the choice of assigning a value from the issue of attention especially if these emotions are linked to evolutionary needs. However it is possible to a certain degree. In particular values which were agreed upon within a community/collective are in some sense subjective, however they are moderated by the collective knowledge. For that reason I would like to call these kind of moderated subjective values *rational subjective values*. Among others they are intended for detaching the subjective emotional evaluation of an individual from highly fluctuating mood changes. An example: If someone stomps on your feet, your reflex could be to stomp back, however usually you would decide to maybe turn away grudgingly because you dont want to behave badly.

A constitution consists to a great part of or is based on rational subjective values. Assertions like wether death penalty is acceptable or not are rational subjective values. There is no scientific method to measure wether killing someone should have the value good or bad. Some communities state that such values were assigned by deities (like e. g. "you shall not kill" in the bible) however they are still something which had been agreed on by a community, so one can keep calling them rational subjective values. Thus an important feature of *rational subjective values is that they are often written down somewhere*.

The above displays in particular that a societal "rule" may be seen as a "value", in that the corresponding society/collective had agreed upon that this rule was "good" or "accepted as a law". Again here the name of a value (namely that what is measured which is: how good do you fulfill a rule) and its actual quantity (quantified value: good) will be for brevity both be called a value.

**side remark:**

*Conversely a measurement can be seen as a rule, in the case of the sieve the rule would be imposed by the physical reality, i. e. the rule could be "you have to fit through the sieveholes" (or the opposite rule not to fit through). Or in other words: by observing wether a sand particle fits through the sieveholes means to observe wether it fulfills the rule "you have to fit through the sieveholes" or not. As a matter of fact: if one accepts this view of a measuring apparatus being a "rule" this could imply that the constancy of a value may be related to the question of "how good the rule" has been "written down". Or in other words: how many similar apparatus' are there/can be built in order to measure a particular value in question. As an illustration: If there is only one particular chunk of a gauge kilogram (which is currently in Paris) and this kilogram is actually changing its size (which it actually does), then the actual value of a particular de facto constant mass is changing accordingly (for that reason physicists are now looking for another standard to measure mass). But if there would have been several identical kilograms from the beginning on then on average this "gauge kilogram decay process" MAY have been slower, i. e. the constancy better. Interestingly this would give a kind*

*of "uncertainty relation" between the rate of change of "on average measurement accuracy" and "space" (i. e. the number of on average ideal copies of measurement apparatus).*

The motivation of a collective to establish rational subjective values is – as already indicated – in order to moderate between the sometimes rather immediate values of the individuum and values which concern the collective, like e. g. how to make a collective survive or to work more efficiently, which often includes nonemotional measurable values such as size of water ressources etc. How efficiently common goals of a collective are achieved is related to the notion of "trust", i. e. to the question of how rules/rational subjective values are accepted in the corresponding collective [RaTru08].

So the term *rational subjective value is always in relation to a collective.*

**rational values, the collective and common goals**

However a collective is usually part of a bigger collective. If the size of the smaller collective is small in comparision to the bigger collective (like a family in a nation) then the values of the smaller collective shall for simplicity also be regarded as individual subjective with respect to the bigger collective. So families or companies may be regarded as individua in relation to a nation, i. e. they are likely to adopt the nations values. If the groups are rather bigger (like e. g. migrants) than their values usually have to be regarded seperately, but this can also be done with respect to a nations values, i. e. they may add different values or reevaluate a nations rational subjective values. Rational subjective values can thus be seen as a kind of average value, averaged in terms of population and time, but still belonging to the individuum (like in terms of a nation these could be called cultural values, in terms of a religious group, religious values, in terms of consumers, consumer values etc.)

Under generic conditions one can assume that a collective behaves mostly according to the given physical circumstances (measurable) and their collective desicions (based on the rational subjective values, laws, rules). However as pointed out above this behaviour is blurred, disturbed by the subjective individual values, depending on the size and nature of societal/collective control.

A collective chooses their values often in order to achieve a set of common collective goals. E. g. the rational subjective value of some christians to explain christianity, to evangelize is good may lead to the common collective christian goal to make everybody turn into a christian.

An important point here is that it is somewhat possible to determine to what extend a choice of rational subjective values meets a common collective goal or in other words: a collective rational subjective value. This can be done by observation. Another example: if a society holds the rational subjective value: killing is not good (you shall not kill) and controls this rational subjective value e. g. via law enforcement then one can observe wether this reduces the overall killing rate (a collective rational subjective value) in that society.

Often collective rational subjective values compete with each other so it is usually never possible to reach one common goal. (like if a religious group decides to advertize their religion in a nonforceful obtrusive manner, as their common goal is not to force people into their believe then this could mean that they may gain less followers). So in short: rational subjective values are somewhat intended to meet collective rational subjective values in an optimized way.

### (computer) models based on values

Interestingly two important physical quantities enter here again, namely time and space. Space enters via the questions of how the individua are connected and having their values compete with each other (the connectivity and the amount of individua has a geographical/spatial component to it), whereas time enters in that it is needed for observing the collective process, which includes the formation of rational subjective values and likewise the observation wether rational subjective values are in accordance with e. g. a collective rational subjective value.

An important fact is that the duration and size of the involved time and space can be abstractly reduced by mathematical computations which model the involved system including its values and the procedural connections (dependencies) between these values. But of course the mathematical model can only approximate the real outcome.

For the accuracy of a mathematical prediction the correct assessment of all individual values as well as all measurable values and their interplay is essential.

For large collections of values this usually turns out to be feasible if one restricts the sample size and identifies the main modes of interaction. Lets call this a choice of sample.

So the accuracy of the mathematical prediction relies to a great part on this choice of sample and the stability of that choice over time.

Roughly one can say that this *mathematical choice of sample is easier and more likely to be accurate if the involved values and their interplay can be measured or* when they are fixed in laws/rules etc. i. e. *if they are rational subjective values.*

This is why complicated physical predictions (like about climate change)

can be quite accurate (as they include mostly measurable values), whereas predictions which involve subjective values, like this is often the case for economic values are usually not so accurate.

I. e. the more an evaluation is less rational (in the above sense) like in panic buyouts at stock markets or like in the recently observed desaster at the societe generale, the harder it is to determine the main involved values and their interdependencies.

In economy the evaluation (the assignment of values) is usually done via pricing (I will call incentives such as interest rates etc. and other economic judgements also pricing in order to make things simpler). So for example the price of a resource depends on the amount of the physically accessible part of that ressource (measureable), the mining regulations in a country (rational subjective values) and e.g. the political situation (these are subjective values as they imply that no collective agreement on values had been achieved).

Lets look at another example, which – due to recent events (the moving of a Nokia plant from Germany to Romania) – may be worthwhile to investigate, namely the value of labour. In a collective, like a nation the value of labour depends among others on

- skill (time and quality of education needed for that skill, amount of people which are in principle capable of acquiring that skill) [more or less measurable value]

- educational resources (amount of capable people who can be trained) [measurable value]

- basic living costs (relative to given prices, like to a basket of available commodities (Warenkorb)) [measurable values in relation to the basket]

- geographical flexibility (this applies to employer as well as to employee: a farmer can hardly move, if his soil is under drought, whereas an agent in a Call Center is not bound to location, however this flexibility may depend also on rational subjective values such as language (journals, journalists, lawyers etc.) and immigration/principal deployment laws. The geographical flexibility also determines how much labour costs are in dependencies to labour costs in other countries. [depending on rational subjective and subjective values]

- negotiation (the prices are under negotiation between the employer and the employee. These depend on cultural (like gender and ethnicity biases) and political powers (trade unions, exclusivity of skill etc.)) [mostly depending on rational subjective values]

- political (national and international politics may subsidize certain sectors, overall conditions may be unstable (which secretly enhances costs) etc. [dependent on rational subjective and subjective values]

The above example illustrates again how mostly subjective values make the pricing into a complicated issue.

Consequently failures in economic predictions are most likely less due to overworked socalled "quants" (mathematicians/physicists in finance) but more likely due to the fact that economy involves non-rational judgements/values.

(As a side remark: It is illustrative to discuss the NOKIA example based on these criteria, since it may display that romanian workers are not necessarily working harder than germans for the money, but that mostly the living conditions are partially cheaper and that the social costs/investments are lower. In particular it is probable that the working conditions (work hours, vacation time) are not as good as in Germany (part of negotiation), however this point most likely doesn't account for the vast differences in wage. Moreover it displays that people with the skills required for working in that NOKIA facility are relatively easy to find. In addition due to high moving costs, immigration, language, cultural and social conditions (and social problems) and to the probably a little smaller Warenkorb (less goods can be purchased for the wage) presumably not too many germans will move to Romania in order to keep working for NOKIA.)

Usually the above kinds of analysis' are done by e. g. traders/economical analysts. Economical decisions are made upon their judgement and ability to assign prices. This is also why they are usually rather well paid. This fact by the way holds true since the beginning of trade. A good trader was able to (more or less) correctly determine prices (usually assigned to a geographical location) in order to know where to buy and where to sell. Of course part of the job of a sales/trade person is also negotiation, but this changes a price usually only partially – the main point is the ability to correctly determine and analyse the prices (last not least the negotiation itself depends on this ability).

**value types and the paradox of value**

It is illustrative to discuss the notion of utility (i. e. the increase of positive values and the decrease of negative values) in terms of the above introduced types of values. Like for example the socalled diamond-water paradox (also called paradox of value), which basically states that it is absurd that the price of diamond is so much higher then the price of water, becomes clearer if one acknowledges the following: Let us assume that there is enough abundance of water for each individuum in a collective (since scarcity could rise the prise of water into arbitrary heights, as it is a necessity for living beings) and that there is no other use for diamond than being used as a jewelry item.

Then the utility of diamond could be very different for the various individua in that collective. In particular some people may even dislike a diamond's color and shape etc. Hence the subjective values/prices for diamonds could apriori greatly differ depending on the individuum. However if a greater share of that collective agrees that say the nice refraction properties of diamond are something that has a positive value (and thus a higher utility) then this implies that the rational subjective value/price of diamond will be higher. This rational subjective value may be related to individual sentiences like that the refraction gives a comforting feeling etc. but it need not to be related apriori. This is also where branding becomes important in that it influences the formation of rational subjective values. Scarcity may amplify a collective agreement on price. Like if it is easy to produce diamond in great amounts then the price would probably approximately be the production/distribution costs, but as it is scarce the price is higher.

On the other hand if a collective agrees that diamond has no or small value, then probably only a few people would dig for it. This is more or less also the reason why the europeans who entered America traded gems against glass perls: the rational subjective values of Americas natives regarding the value of glass versus gem where different then the ones of European natives, which was of course partially due to the fact that Americas natives didnt know about glass production.

It should also be mentioned that besides being a jewelry item diamond usually includes of course other uses. In particular due to its scarcity, solidity and recognizability/recall value it may serve as medium of exchange.

**value assignment and tradability**

The correct assignment of prices can get arbitrarily complicated and so mathematicians use for that problem awkward sounding tools like risk calculus of von Neumann - Morgenstern or Arrow-Pratt risk aversion. The above should thus only serve as an introductory exposition how science enters societal and economical questions.

There exists various platforms for financial analysis' like e. g. Bloomberg L.P. which provide background information (including political) and software tools. Likewise professional analysts and e. g. rating agencies are adding their expertise to the financial and economical world.

So why should a scientific platform provide a similar service?

As was already pointed out before a lot of the pricing mechanism depends on the ability of traders to analyse and judge about prices. Here a funny feedback happens. In particular the accuracy of pricing depends also on the property of being tradeable. Or in other words if there is no trading reason for assigning a price to something then this missing price will destort the actual price. This explains why such subjective values, which are often even rational subjective values for nations (!) like a healthy environment for future generations or human social conditions are not necessarily rational subjective values for traders. This implies that the pitfalls of capitalism are also due to that organisatorial problem of an inaccurate and/or missing pricing of values beyond the trading scope. It makes the thinking about capitalism less into an ideological question.

However it is of course possible to assign values/prices beyond the trading scope. This has e. g. been demonstrated in the socalled Stern review, where the future costs of climate change were estimated. However these prices are only sparsely reflected in nowadays economy and trade.

Similar estimations could be made for nuclear energy. Nuclear energy is currently (especially with respect to regenerative energies) relatively cheap, however judged by environmental costs and future environmental costs (like over the next 20 000 years) it is very expensive [this is a measurable value]. So the actual traded price for nuclear energy is simply one thing: wrong.

Likewise the almost "zero value" of digitized items, like e. g. the one of digital music are in fact almost zero value only in terms of trading. It is not true that these items have no value or no costs. Every musician would agree on this.

Thus it is not too far fetched to conjecture that these items are not so interesting for the trading market, because the *value exchange* for digital items , i. e. the *trading* of digital items is – due to e. g. private copying – almost non-manageable. This holds also true for e.g. digital storage space. Here the actual costs are usually acclaimed somewhere else, as was already pointed out by Chris Anderson [And08] which he - among others - illustrated at the example of Google. This mechanism is a bit comparable to dumping prices, just that dumping prices need not to be due to a non-managibility of the trading process.

As a matter of fact non-tradability takes also place in the case of a monopole, where the actual price and the real costs are often in a misrelation.

Concluding: There is a need to analyse the price of goods also with respect to apriori nontradable values, like pollution, social conditions, future living conditions etc. As already explained an economic platform will do this only to a limited extend. Institutes like the New Economics Foundation [NEF] have usually only rather limited resources to provide such a service. Small institutions may also be less neutral.

A global platform, where economy/financial mathematics departments of universities could bring in their expertise, where agriculture experts, climate scientists, cultural experts etc. work together would have a different quality.

As a side effect such a service could even have an economic value, as e. g. insurance companies often need to find substitutes for these pricing informations in things like future risks. This may encourage on the other hand the financial world to donate financial information to the proposed internet platform.

Last not least a more detailled and true analysis of prices would allow for political measurements similar to the emission tradings. One could e. g. introduce a similar thing for the future costs of nuclear waste, for deforestation etc. One could introduce certificates for maximum work hours, health benefits to employees etc. (although legal regulations are likely to be more efficient in some of these cases). A more correct and scientifically sound determination of a price is a necessary condition for that.

As already indicated above the actual "how" of the determination of a value is very complicated and beyond the scope of this article. Often it is already helpful to understand which main pricing processes are involved and how they depend on each other. The notion "determination of a price" certainly includes "fuzzy" determinations. Or in other words it is already helpful if a big round of experts could assess economical/societal etc. price ranges and approximate future developments and thus assist in making complicated political decisions, like e. g. in the case of mandated markets.

A scientifically minded determination of a price could eventually also include game-like structures like betting on prices, a kind of toy stockmarket etc. Also if I do not agree with Robin Hanson that "Betting markets are our best known institution for aggregating information." [Ha] behavourial strategies like bargaining, bluffing, risk aversion etc. (see e.g. [Ca03]) play an important role in economics and last not least in the determination of a price.

# 4 Acknowledgements

I would like to thank Sabine Hossenfelder for valuable help. Our communication [Bee08] can be found in the comment section to a post on the blog randform on which I prepublished most of the thoughts, which went into this article. It is also on her blog Backreaction, where I found the link to senseaboutscience [SAS].

I found the link to Robin Hansons article [Ha] on Scott Aaronsons blog.[Aa]
I would like to thank Tim Hoffmann for a critical reading of the document.